\documentclass[aps,prc,letterpaper,superscriptaddress,showpacs,amsmath,amssymb,nofootinbib,twocolumn]{revtex4-1}

\usepackage{graphicx}  
\graphicspath{ {figures/} } 
\usepackage{bm}     
\usepackage{amssymb}   
\usepackage{amsmath}   
\usepackage{hyperref}
\usepackage{todonotes}
\usepackage[normalem]{ulem}
\usepackage[utf8]{inputenc}
\usepackage[acronym,style=super,nonumberlist,toc]{glossaries} 
\usepackage{xcolor}

\newcommand\extrafootertext[1]{
    \bgroup
    \renewcommand\thefootnote{\fnsymbol{footnote}}
    \renewcommand\thempfootnote{\fnsymbol{mpfootnote}}
    \footnotetext[0]{#1}
    \egroup
}

\begin{document}
\newcommand{\ornlphys}{Physics Division, Oak Ridge National Laboratory, Oak Ridge, TN 37831, USA}
\newcommand{\ornlneutrontech}{Neutron Technologies Division, Oak Ridge National Laboratory, Oak Ridge, TN 37831, USA}
\newcommand{\ncsu}{Department of Physics, North Carolina State University, Raleigh, NC 27695, USA}
\newcommand{\tunl}{Triangle Universities Laboratory, Duke University, Durham, NC 27708, USA}
\newcommand{\uky}{Department of Physics and Astronomy, University of Kentucky, Lexington, KY 40506, USA}
\newcommand{\utk}{Department of Physics and Astronomy, University of Tennessee, Knoxville, TN 37996, USA}
\newcommand{\uva}{Department of Physics, University of Virginia, Charlottesville, VA 22904, USA}
\newcommand{\asu}{Department of Physics, Arizona State University, Tempe, AZ 85287, USA}
\newcommand{\mani}{Department of Physics, University of Manitoba, Winnipeg, MB R3T 2N2, Canada}
\newcommand{\umich}{Department of Physics, University of Michigan, Ann Arbor, MI 48109, USA}
\newcommand{\utc}{Department of Physics and Astronomy, University of Tennessee at Chattanooga, Chattanooga, TN 37403, USA}
\newcommand{\lanl}{Los Alamos National Laboratory, Los Alamos, NM 87545, USA}

\title{First Full Dalitz Plot Measurement in Neutron \texorpdfstring{$\beta$}{Beta}-Decay using the Nab Spectrometer and Implications for New Physics} 

\author{Francisco M. Gonzalez}
\email[Email: ]{gonzalezfm@ornl.gov}
\affiliation{\ornlphys}
\author{Jin Ha Choi}
\affiliation{\ncsu}
\affiliation{\tunl}
\author{Himal Acharya}
\affiliation{\uky}
\author{Skylar Clymer}
\affiliation{\asu}
\author{Andrew Hagemeier}
\affiliation{\uva}
\author{David G. Mathews}
\affiliation{\ornlphys}
\affiliation{\uky}
\author{August Mendelsohn}
\affiliation{\mani}
\author{Austin Nelsen}
\affiliation{\uky}
\author{Hitesh Rahangdale}
\affiliation{\utk}
\author{Love Richburg}
\affiliation{\utk}

\author{Ricardo Alarcon}
\affiliation{\asu}
\author{Ariella Atencio}
\affiliation{Department of Physics, Drexel University, Philadelphia, PA 19104, USA}
\author{Stefan Bae{\ss}ler}
\affiliation{\uva}
\affiliation{\ornlphys}
\author{Thomas Bailey}
\affiliation{\ncsu} \affiliation{\tunl}
\author{Noah Birge}
\affiliation{\utk }
\author{Dennis Borissenko}
\affiliation{\uva}
\author{Michael Bowler}
\affiliation{\uva}
\author{Leah J. Broussard}
\affiliation{\ornlphys}
\author{Albert T. Bryant}
\affiliation{\uva}
\author{Jimmy Caylor}
\affiliation{\utk }
\author{Tim Chupp}
\affiliation{\umich}
\author{Christopher Crawford}
\affiliation{\uky}
\author{R. Alston Croley}
\affiliation{\uky}
\author{Micah Cruz}
\affiliation{\utk }
\author{George Dodson}
\affiliation{Massachusetts Institute of Technology, Cambridge, MA 02139, USA}
\author{Wenjiang Fan}
\affiliation{\uva}
\author{Deion Fellers}
\affiliation{Lawrence Berkeley National Laboratory, Berkeley, CA 94720, USA}
\author{Nadia Fomin}
\affiliation{\utk}
\author{Emil Frle\v{z}}
\affiliation{\uva}
\author{Matthew Frost}
\affiliation{\ornlneutrontech}
\author{Jason Fry}
\affiliation{Department of Physics, Geosciences, and Astronomy, Eastern Kentucky University, Richmond, KY 40475, USA}
\author{Duncan Fuehne}
\affiliation{Department of Physics, University of Chicago, Chicago, IL 60637, USA}
\author{Michael T. Gericke}
\affiliation{\mani}
\author{Michelle Gervais}
\affiliation{\uky}
\author{Corey Gilbert}
\affiliation{\ornlphys}
\author{Ferenc Gl\"{u}ck}
\affiliation{Karlsruhe Institute of Technology, IAP, 76021 Karlsruhe, POB 3640, Germany.}
\author{Rebecca Godri}
\affiliation{\utk}
\author{Geoff L. Greene}
\affiliation{\utk}
\affiliation{\ornlphys}
\author{William Greene}
\affiliation{\utk}
\author{Josh Hamblen}
\affiliation{\utc}
\author{Paul Harmston}
\altaffiliation[Now at: ]{ Department of Physics, University of Illinois at Urbana-Champaign, Urbana IL, 61801, USA}
\affiliation{\utk} 
\author{Leendert Hayen}
\affiliation{LPC Caen, ENSICAEN, Université de Caen, CNRS/IN2P3, Caen, France}
\affiliation{\ncsu}
\author{Carter Hedinger}
\affiliation{\uva}
\author{Chelsea Hendrus}
\affiliation{\umich}
\author{Sean Hollander}
\affiliation{\ornlphys}
\author{Kavish Imam}
\affiliation{\utk}
\author{Erik B. Iverson}
\affiliation{\ornlneutrontech}
\author{Aaron Jezghani}
\affiliation{Partnership for an Advanced Computing Environment, Georgia Tech, Atlanta, GA  30332, USA}
\author{Chenyang Jiang}
\affiliation{\ornlneutrontech}
\author{Huangxing Li}
\affiliation{\uva}
\author{Nick Macsai}
\affiliation{\mani}
\author{Mark Makela}
\affiliation{\lanl}
\author{Russell Mammei}
\affiliation{Department of Physics, University of Winnipeg, Winnipeg, MB R3B 2E9, Canada}
\author{Ricky Marshall}
\affiliation{\uva}
\author{Madelyn Martinez}
\affiliation{\asu}
\author{Mark McCrea}
\affiliation{\uky}
\author{Pat McGaughey}
\affiliation{\lanl}
\author{Sean McGovern}
\affiliation{\uva}
\author{David McLaughlin}
\affiliation{\uva}
\author{Jacqueline Mirabal-Martinez}
\affiliation{\lanl}
\author{Paul Mueller}
\affiliation{\ornlphys}
\author{Andrew Mullins}
\affiliation{\uky}
\author{William Musk}
\affiliation{\uva}
\author{Jordan O'Kronley}
\affiliation{\utk}
\author{Seppo I. Penttil\"{a}}
\affiliation{\ornlphys}
\author{D. Elliot Perryman}
\altaffiliation[Now at: ]{Institut Laue-Langevin, Grenoble, France}
\affiliation{\utk}
\author{Josh Pierce}
\affiliation{Neutron Scattering Division, Oak Ridge National Laboratory, Oak Ridge, TN 37831, USA}
\author{Jason A. Pioquinto}
\affiliation{\uva}
\author{Dinko Po\v{c}ani\'{c}}
\affiliation{\uva}
\author{Hunter Presley}
\affiliation{\utk}
\author{John Ramsey}
\affiliation{\ornlphys}
\author{Glenn Randall}
\affiliation{\asu}
\author{Zachary Raney}
\affiliation{\uva}
\author{Jackson Ricketts}
\affiliation{\utc}
\author{Grant Riley}
\affiliation{\lanl}
\affiliation{\utk}
\author{Americo Salas-Bacci}
\affiliation{\uva}
\author{Sepehr Samiei}
\affiliation{\uva}
\author{Alexander Saunders}
\affiliation{\ornlphys}
\author{Wolfgang Schreyer}
\affiliation{\ornlphys}
\author{E. Mae Scott}
\altaffiliation[Now at: ]{Centre College, 600 West Walnut Street, Danville, KY 40422}
\affiliation{\utk} 
\author{Thomas Shelton}
\affiliation{\uky}
\author{Aryaman Singh}
\affiliation{\uva}
\author{Alexander Smith}
\affiliation{\uva}
\author{Erick Smith}
\affiliation{\lanl}
\author{Eric Stevens}
\affiliation{\uva}
\author{R. J. Taylor}
\affiliation{\ncsu} \affiliation{\tunl}
\author{Leonard Tinius}
\affiliation{\uva}
\author{Isaiah Wallace}
\affiliation{\utk}
\author{Jonathan Wexler}
\affiliation{\ncsu} \affiliation{\tunl}
\author{W. Scott Wilburn}
\affiliation{\lanl}
\author{A. R. Young}
\affiliation{\ncsu} \affiliation{\tunl}
\author{B. Zeck}
\affiliation{\ncsu} \affiliation{\tunl}
\collaboration{Nab Collaboration}

\begin{abstract}

Precision measurements of observables in neutron $\beta$-decay are used to test the Standard Model description of the weak interaction and search for evidence of new physics. 
The Nab experiment at the Fundamental Neutron Physics Beamline at the Spallation Neutron Source was constructed to measure correlations in neutron decay by utilizing an asymmetric spectrometer and novel detection system to accurately reconstruct the proton momentum
and electron energy for each $\beta$-decay. 
This work describes the detection of neutron $\beta$-decay products in the Nab spectrometer and presents the first full Dalitz plot representation of the phase space of neutron $\beta$-decay for all electrons $>100$~keV. 
In addition, new constraints are placed on a possible excited neutron state, hypothesized to explain the disagreement between the appearance and disappearance neutron lifetime techniques.

\end{abstract}

\maketitle
\extrafootertext{Notice: This manuscript has been authored by UT-Battelle, LLC, under contract DE-AC05-00OR22725 with the US Department of Energy (DOE). The US government retains and the publisher, by accepting the article for publication, acknowledges that the US government retains a nonexclusive, paid-up, irrevocable, worldwide license to publish or reproduce the published form of this manuscript, or allow others to do so, for US government purposes. DOE will provide public access to these results of federally sponsored research in accordance with the DOE Public Access Plan (https://www.energy.gov/doe-public-access-plan).}

\section{Introduction} 
\label{intro}

Neutron $\beta$-decay, the simplest example of nuclear $\beta$-decay, provides a unique suite of precise tests of the Standard Model (SM) of particle and nuclear physics. 
An important constraint of our understanding of the weak interaction is the unitarity of the quark-mixing Cabibbo-Kobayashi-Maskawa (CKM) matrix, with the most precise test of unitarity utilizing the first row of matrix elements. 
Recent progress in calculating the radiative corrections in nuclear decays have led to re-evaluations in the CKM matrix element $V_{ud}$ extracted from superallowed $0^+\rightarrow0^+$ Fermi nuclear decays\cite{Seng:2018qru,Gorchtein:2018fxl,Hardy:2020qwl}. 
This has led to significant tension between the SM prediction of CKM unitarity and the leading experimental extractions of the CKM matrix elements $V_{ud}$ and $V_{us}$ used in the first-row test~\cite{FlavourLatticeAveragingGroup:2019iem,Falkowski:2020pma}.
Resolving this anomaly, also known as the Cabbibo Angle Anomaly, has implications for physics Beyond the Standard Model (BSM)~\cite{Cirigliano:2023nol}, 
including
right-handed hadronic currents~\cite{Cirigliano:2022yyo}, lepton flavor universality~\cite{Crivellin:2020lzu}, and the $W$-boson mass~\cite{Cirigliano:2022qdm}. 

Neutron $\beta$-decay provides a complementary method of extracting the $V_{ud}$ matrix element, free from nuclear-structure-dependent theoretical corrections that dominate the uncertainty when using superallowed Fermi nuclear decays. 
Two neutron-decay parameters need to be determined more precisely to directly probe the Cabbibo Angle Anomaly: the neutron lifetime $\tau_n$ and the ratio of the Gamow-Teller and Fermi coupling constants $\lambda = g_A / g_V$. 

While recent measurements of the neutron lifetime approach the required precision to compete with $0^+\rightarrow0^+$ decays~\cite{UCNt:2021pcg, 
Musedinovic:2024gms},
there is a 
more than $4\sigma$ discrepancy between experiments that measure the disappearance of stored neutrons and the appearance of decay products~\cite{ParticleDataGroup:2024cfk}. 
This discrepancy has motivated exotic explanations, such as dark decays~\cite{Fornal:2018eol}, mirror matter~\cite{Berezhiani:2018eds, Berezhiani:2018qqw, Tan:2023mpj}, excited state neutrons~\cite{Koch:2024cfy}, and a second flavor of atomic hydrogen~\cite{Oks:2025gkl}. 
Experimental searches for these exotic decays, species, and transitions have been ongoing but thus far have not seen evidence for new physics~\cite{Tang:2018eln,UCNA:2018hup,Klopf:2019afh,Broussard:2021eyr,LeJoubioux:2023usk}, and exotic solutions to the neutron lifetime discrepancy are constrained by the consistency of 
$V_{ud}$ as extracted from neutron $\beta$-decay 
and nuclear $\beta$-decay~\cite{Dubbers:2018kgh}.

The other free parameter in neutron decay, $\lambda$, can be 
extracted by measuring correlations in neutron decay. The $\lambda$ parameter has been determined most precisely from the asymmetry $A$~\cite{Markisch:2018ndu,UCNA:2017obv}, which correlates the momentum of the produced electron with the spin of the decaying neutron, and from the correlation parameter $a$~\cite{Hassan:2020hrj,Beck:2019xye,Beck:2023hnt,ACORN:2024}, which correlates the momenta of the electron and neutrino produced in the decay. 
Currently, there is some tension between the measurements of $\lambda$ using the two techniques~\cite{ParticleDataGroup:2024cfk}. 
As the precision in calculations of the axial vector coupling constant $g_A$~\cite{Chang:2018uxx, Ma:2023kfr} and understanding of relevant corrections improve~\cite{Cirigliano:2022hob, Seng:2024ker, Cirigliano:2024nfi}, directly comparing determinations of $\lambda$ experimental results with improved lattice calculations can also be used to constrain BSM physics. 

Beyond extracting $V_{ud}$, precision nuclear and neutron decay experiments can probe exotic couplings in the weak interaction~\cite{QuarkLevelCouplingsCirigliano:2012,Gonzalez-Alonso:2018omy}. 
The existence of a non-zero Fierz interference term $b$ would be indicative of exotic, possibly right-handed, scalar or tensor couplings beyond the known $V-A$ structure of the weak interaction. 
This term can be measured precisely in $\beta$-decays, including that of the neutron, with sensitivity comparable to high-energy limits established at the Large Hadron Collider~\cite{Falkowski:2020pma}. Scalar currents are presently most tightly constrained by the set of superallowed Fermi decays~\cite{Hardy:2020qwl}. 
Tensor currents have recently been precisely probed in nuclei such as $^{8}$Li and $^{8}$B~\cite{Burkey:2022gpb,Longfellow:2024svx}, and future precision studies will use cyclotron radiation emission spectroscopy with $^6$He and $^{19}$Ne~\cite{King:2022zkz,He6-CRES:2022lev}.
Neutron correlation experiments 
have also placed limits on $b$ in neutron decay competitive with nuclei~\cite{Sun:2019dlk,Saul:2019qnp,Beck:2023hnt}.
Recently, a combined analysis from two neutron decay correlation experiments measuring different asymmetry parameters found a non-zero value of $b$~\cite{Beck:2023hnt}, deviating from the 
SM prediction by $2.8\sigma$. A non-zero value of $b$ of that size had been proposed earlier as part of a resolution of the neutron lifetime discrepancy involving dark decays~\cite{Ivanov:2018uuk}.

\begin{figure}[h]
    \centering
    \includegraphics[width=\columnwidth]{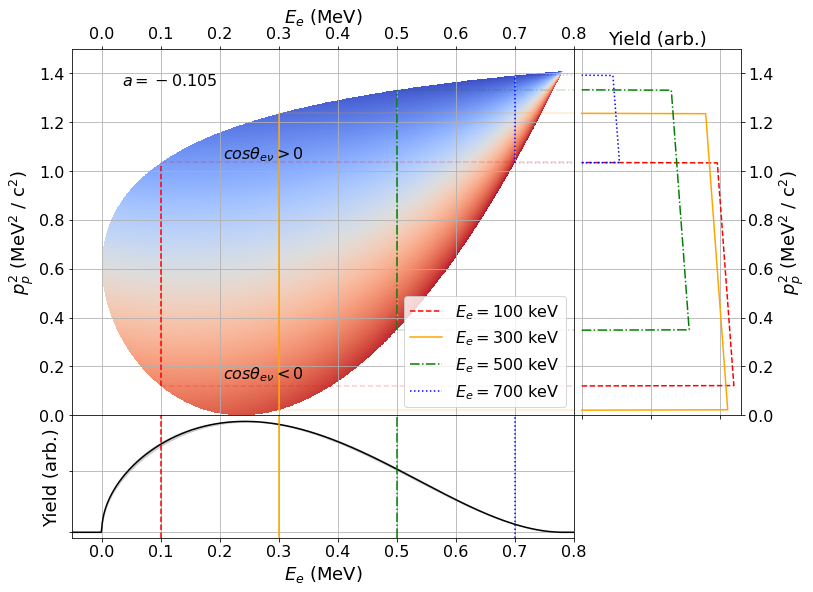}
    \caption{Analytical calculation, neglecting radiative and recoil corrections, showing the phase space of neutron decay kinematics as a function of electron kinetic energy $E_e$ and proton momentum $p_p$. The central plot shows the available phase space. The blue space represents decays with the electron and antineutrino momenta aligned within the same hemisphere, and the red space represents decays with the electron and antineutrino momenta antialigned. The right plot shows the relative proton yield for slices of constant $E_e$. The lower plot shows the overall electron distribution summing over all values of $p_p^2$. (Color online)}
    \label{fig:intro:kinematics}
\end{figure}

Rather than relying on detecting the antineutrino, 
the neutron decay correlation parameters can be reframed in terms of the more experimentally accessible decay electron and proton energies. 
The values of $a$ and $b$ can be then determined by measuring the energy of the electron 
produced in $\beta$-decay and the momentum of its coincident proton, 
with the kinematically available phase space seen in Fig. ~\ref{fig:intro:kinematics}. 
This plot is analogous to the commonly used Dalitz plot in high-energy particle physics for reconstructing three-body kinematics and looking for resonant decay modes~\cite{Dalitz:1953cp}. 
The proton yield for slices of constant electron kinetic energy $E_e$ has a slope proportional to $a$, and the proton momenta of its sharp edges are determined entirely by the decay kinematics. Changes in $b$ will cause a distortion of the electron energy spectrum.
Exotic decays may appear as deviations from the kinematically allowed edges of the Dalitz plot.

As previous neutron $\beta$-decay experiments have not measured the complete range of $p_p$ and $E_e$ simultaneously, the full Dalitz plot of neutron $\beta$-decay has not yet been experimentally reconstructed. 
Experiments measuring correlations with the neutron spin typically precisely measure only $E_e$~\cite{UCNA:2017obv,Markisch:2018ndu}. 
Measurements of the electron-neutrino correlation, $a$, measured the proton recoil spectrum~\cite{Beck:2019xye} or narrow slices of phase space~\cite{Hassan:2020hrj}. 
Other neutron decay searches have measured proton kinematics by using the time-of-flight between protons and electrons, but these have not provided a precise measurement of the electron energy or proton momentum~\cite{Chupp:2012ta, Broussard:2016gqg}. 
In this work, we present a first reconstruction of the full neutron $\beta$-decay Dalitz plot above thresholds and place constraints on possible excited decay modes of the neutron. 

\section{Experimental Apparatus}
\label{sec:apparatus}

\begin{figure}
    \centering
    \includegraphics[width=\columnwidth]{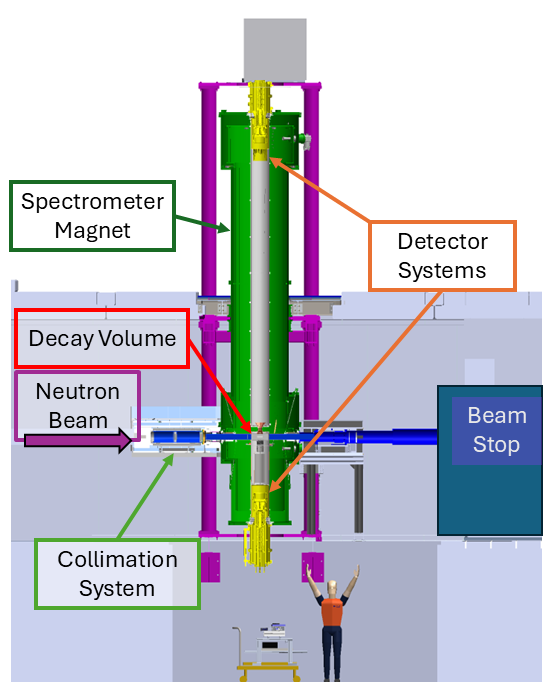}
    \caption{Elevation view of the Nab experiment showing key subsystems, including the beamline, magnet, and detector systems. Detailed descriptions of the subsystems are provided in the text.}
    \label{fig:setup}
\end{figure}

The measurement reported here utilizes the Nab spectrometer, located at the Spallation Neutron Source (SNS) at Oak Ridge National Laboratory, designed to accept the full phase space of neutron $\beta$-decay. 
The Nab apparatus was developed to determine $\lambda$ with high precision through a measurement of the electron-neutrino correlation, $a$, and set strong constraints on a possible Fierz term, $b$~\cite{NabBowman:2005, Nab:2008cwh, Baessler:2014gia, Fry:2018kvq}. 
To measure $E_e$ and $p_p$ for each decay event, the spectrometer's magnetic field captures the charged decay products from the unpolarized neutrons and guides them to two detection systems. 
After the fast decay electron is detected and its energy measured directly, the lower-velocity coincident proton has to travel along the long arm of the spectrometer before being detected several microseconds later. 
This time of flight determines the proton momentum.
An overview of the apparatus is depicted in Fig.~\ref{fig:setup}, and major components as implemented for this work are described in more detail below. 

\subsection{Neutron Beam, Spectrometer, and Magnet}

The Nab apparatus is installed at the Fundamental Neutron Physics Beamline 13-B (FNPB)~\cite{Fomin:2014hja} of the SNS. 
Neutrons were produced via spallation by a 60\,Hz, $1.7$~MW proton beam impinging upon a mercury target. 
The beamline monochromator was inserted during data-taking for these measurements to share beam with experiments on the adjacent Beamline 13-A, resulting in a modest reduction in flux. 
At the exit of the biological shielding of the FNPB, the Nab beamline consisted of an upstream section, the path through the spectrometer, a downstream section, and the neutron beam stop.
For this measurement, the upstream section included a spin flipper~\cite{Hendrus:2023the} which was not operated, but provided a vacuum environment for the neutrons to reduce neutron-generated background radiation. 
Thin $0.3$\,mm magnesium-alloy windows separate the upstream and downstream rough beamline vacuum from the spectrometer's ultra-high vacuum, with a residual gas pressure below $10^{-8}$\,Torr. 
Inside the spectrometer magnet\footnote{\url{https://www.cryogenic.co.uk/news-and-events/news/actively-shielded-nab-spectrometer}}, the beam is surrounded by $^6$Li-loaded glass channels upstream and downstream of the decay volume. Additional collimators made of $^6$Li$_3$PO$_4$ are mounted in the upstream section. The collimators nominally result in a rectangular beam with a width of $5.4$~cm and a height of $7$~cm in the decay volume. 
Upon exiting the spectrometer, the beam passes a second magnesium-alloy window and follows an evacuated vacuum tube until it is absorbed in a shielded enclosure containing the neutron beam stop.  

A tiny fraction, less than $10^{-6}$, of the total neutrons passing through the spectrometer decay inside of a small decay volume within the $\sim1.7$\,T magnetic field of the spectrometer. 
The charged decay products gyrate inside this magnetic field and are guided by it to silicon detectors at either end.
The magnetic field of the spectrometer and the agreement of its measurement with calculation were described in Ref.~\cite{Fry:2018kvq}. 
Above the decay volume, the magnetic field strength increases to $\sim4$\,T to reject protons emitted at an angle far from the spectrometer axis which would take a long time to reach the detector. 
This produces a sensitive response function between the proton time of flight $t_{p}$ and its momentum $p_p$. 
Beyond this pinch, the field drops to a low value of $\sim0.2$\,T to longitudinalize the momentum of the protons. 
At either end of the spectrometer, the magnetic field again rises to $\sim1.3$\,T to guide the charged particles into the detector system.

Since the mass of the proton is significantly greater than the mass of the electron, conservation of energy and momentum, combined with the more than $5$~m path length of the proton, causes at least a $12$~$\mu$s difference in the time of flight between the proton and electron.
Since we do not observe the initial conditions (e.g. position or momentum) of the neutron decay, we must rely on coincidences between protons and electrons close to where the same field lines intersect the detector. 
The guiding fields between detectors ensure that the electron and proton from a given decay hit the detectors within $\sim2$\,cm of one another. 
In this work, we define the time of flight $t_{pe}$ as the time difference between electron and proton triggers.

\subsection{Detection System}
\label{ssec:silicon}
The goal of the detection system is to detect the decay proton with up to $0.751$~keV and the decay electron with up to $782$~keV in coincidence, whether in the same or different detectors. 
The requirements for the system include: low noise; a dead layer thin enough to detect the protons after accelerating them to no more than 30~keV; excellent linearity and resolution for electron energy reconstruction; fast timing to resolve backscattered electrons and reduce biases in reconstructing the proton time of flight; and high segmentation to reduce noise and backgrounds in coincidence detection. The detection system is based on thick, large-area, highly segmented silicon detectors with custom-designed electronics and a compact assembly providing electrical power, data acquisition, and cryogenic cooling while floating on a $-30$~kV accelerating potential. 
Two detector systems are used, one at each end of the Nab spectrometer. 
The performance of a prototype detector system has been demonstrated in~\cite{Broussard:2016gqg} and the final design with full instrumentation has been described in Ref.~\cite{Broussard:2017tab}. 

This measurement used 2\,mm thick silicon detectors with 11\,cm diameter and an active area segmented into 127~hexagons, each with 70\,mm$^2$ area, manufactured by Micron Semiconductor Ltd.\footnote{\url{https://www.micronsemiconductor.co.uk/}} A detailed charge transport simulation model of the detector has been developed~\cite{Hayen:2022dsg}. The detector thickness ensures the highest energy electrons from neutron $\beta$-decay and from calibration sources are fully stopped in the detector. 
These detectors have been demonstrated to have a $\sim$100\,nm dead layer, sufficiently thin to allow efficient detection of 30\,keV protons~\cite{Salas-Bacci:2014xpo}. The upper detector used for proton detection in this measurement was successfully tested using a separate 30\,keV proton beam at the University of Manitoba~\cite{Macsai:2023the}. 

For this measurement, the detectors were cooled to temperatures between $110$~K and $150$~K (depending on the study) to reduce dark noise and the detector charge-collection rise time. 
The upper detector was typically operated with a bias voltage exceeding $-200$~V, the nominal voltage to fully deplete the 2\,mm thick bulk silicon. 
On the lower detector, an abnormally high leakage current ($\sim 100$~$\mu$A at $\sim130$~K) and the resulting excessive noise limited the bias voltage to $-50$~V, below full depletion, which has implications for both timing and energy reconstruction. 

Each detector system was supported by a custom electronics package to provide high gain while maintaining the fast timing intrinsic to the silicon detector response, with 127 channels each reading a hexagonal pixel of the detector. The circuit used in this measurement is described in Ref.~\cite{Hayen:2022dsg}. During this measurement, mechanical and electrical failures resulted in a reduction to the number of operating pixels, such that there were no pixels with all adjacent neighbors operational. 

The detector system~\cite{Broussard:2017tab} included the mounting structure for the Si detector, the detector electronics, instrumentation for gaseous helium flow cooling for the Field Effect Transistors and detector~\cite{Richburg:2025nyp}, water cooling to stabilize the temperature of the amplification stages of the electronics~\cite{Nelsen:2024The}, and vacuum services for the cold section of the electronics. 
It also provided for -30\,kV high voltage isolation of the detector, needed to accelerate protons past the detector dead layer. 
The associated detector electronics chain, including the frontend data acquisition (DAQ) system, the detector bias voltage power supply and pulser, and instruments for monitoring and networking, was held at the -30\,kV potential as well. 
During the data acquisition described here, the upper detector was set to $-30$~kV, while the lower detector was grounded. The applied voltage was characterized by studying the energy shift due to the varying voltage on the mono-energetic conversion electron lines from $^{113}$Sn. 
The decay volume in the section of the spectrometer at neutron beam height was grounded by an electrode system to maintain a constant electric field for all decaying particles~\cite{Li:2021the}. In the lower section of the spectrometer, a second electrode system created up to a 1\,kV potential difference perpendicular to the spectrometer magnetic field. This was operated for some datasets to sweep downward-going, wrong-direction protons far enough away from the electrons that they would no longer be considered coincident even if they were backscattered from the lower detector and subsequently detected in the upper one. 

\subsection{Calibration System}
The response of the silicon detector and electronics chain was characterized and the energy response calibrated \textit{in situ} using radioactive sources. 
Conversion electron sources were selected to encompass the range of neutron $\beta$-decay energies. 
These sources included $^{207}$Bi, which has prominent monoenergetic peaks at $481.7$~keV and $975.7$~keV; $^{113}$Sn, which has its most prominent peak at $363.8$~keV; and $^{109}$Cd, with its most prominent peak at $62.5$~keV~\cite{Nudat}. 
$^{148}$Gd, a monoenergetic 3.18\,MeV $\alpha$ emitter, was also used as an \textit{in situ} probe of the silicon dead layer, important in characterizing the energy lost by the decay protons. 

To reduce the energy lost by decay electrons in the radioactive sources themselves, the isotopes were prepared as a liquid solution, applied in a series of microliter droplets onto thin foils, and installed in the spectrometer after the solvent had evaporated.
An electroplated $^{148}$Gd source was used for studies requiring $\alpha$ particles. 
The sources were mounted onto a custom radioactive source insertion system, which allowed sources to be exchanged and inserted into the ultra-high vacuum of the magnet. 
This insertion system was located at the nominal midpoint of the neutron decay volume along the neutron beam travel axis and was retracted when the neutron beam was on.  
For these measurements, the insertion system could only be actuated along an axis perpendicular to the neutron beam. 
The electron sources were used to calibrate individual pixels on the detector by inserting them so that each source illuminated one cluster of about three pixels of each detector at a time. 

\subsection{Data Acquisition and Signal Processing}
\label{ssec:daq}

The data acquisition system used in this measurement was described in detail in Ref.~\cite{Mathews:2024}. 
The signal waveform outputs from the detector electronics were recorded by National Instruments 8-channel 250 MS/s 14-bit PXIe-5171 reconfigurable oscilloscope modules~\cite{PXIe5171}. 
At each detector, $16$ of these modules were installed in a PXIe-1085 chassis~\cite{PXIe1085}. 
Each detector channel was attached to a dedicated Analog-Digital Converter (ADC) channel, such that signals from each detector pixel could be read out individually. 

As part of the DAQ startup, the internal clocks of each module are synchronized with one another. 
During this synchronization routine, ADC modules within one chassis were synchronized at the better than $1$~ns level. Between the two chassis there was a potential timing offset due to how the synchronization logic managed their cable length differences. 
With the firmware used during this dataset, synchronization between the two chassis was found to be inconsistent. Using the observed offsets in the time of flight of backscattered electrons traveling down-to-up and up-to-down compared to simulation, the timing offset averaged over all runs was determined to be $260 \pm 5$~ns where the $\pm 5$~ns uncertainty was dominated by timing drift between individual runs.

Charged particle events are detected by convolving the raw ADC data stream with a double trapezoid response function on each oscilloscope module's FPGA. 
The double trapezoid, which uses a delayed second negative trapezoid in addition to the more typical positive first trapezoid, cancels baseline drifts that might otherwise lead to compromised trigger efficiency. It also provides a zero-crossing, similar to a constant fraction discriminator, for better timing.
The double-trapezoid trigger filter had a rise time of $1.6$~$\mu$s and a flat top of $0.4$~$\mu$s. 
During this dataset, the DAQ applied a threshold with voltage corresponding to approximately $9$\,keV for the upper detector (the only detector able to see protons) and $90$\,keV for the lower detector. 
\section{Experimental Approach}

\subsection{Data Collection} 
\label{sec:data}

Data were collected during commissioning of the experimental apparatus during a period of three weeks in the summer of 2023, with the intent to demonstrate the working principles of the experiment. All major subsystems were operational. 
Approximately $20$\,\% of all data taken during this commissioning period, $101$~hours, were taken in nominal ``production'' mode, while the remainder of time was used to explore variations in operating parameters. 
``Production'' data were selected from measurement runs with the SNS accelerator power at $1.7$~MW, with the magnet running at the designed field, and the upper detector having an accelerating potential of -$30$~kV. 
While data were being recorded, the uptime of neutron data taking was $75$\,\%, with losses primarily due to planned and unplanned accelerator outages, time for reconfiguration of the experiment and DAQ, and periods where increased electronics noise temporarily paused data acquisition. 
Systematic studies explored in this dataset included detector characterizations using radioactive sources, studies of backgrounds, and studies with varying operating parameters of the detectors and data acquisition. 

Several limitations caused deviations from the ideal operating conditions expected for the final implementation of the experiment, summarized here. 
The limited number of operational pixels reduced the observed coincidence rate and also distorted the energy spectrum due to missed backscattered electrons. 
The implemented version of the timing synchronization routine for the DAQ resulted in a larger than expected uncertainty in the timing offset between the clocks of the upper and lower systems. 
The conditions of the neutron beam varied, both due to variations in the moderator and to accommodate other experiments utilizing BL-13A. 
A lower-than-expected proton energy on the upper detector caused reduced trigger efficiency and an increased timing jitter. 
The lower detector could only be operated with partial depletion, which affected the precision of time-of-flight extraction and the energy reconstruction. 
Finally, the detector temperatures and bias voltages were not held in a consistent and stable configuration for long periods during production mode due to emphasis on optimizing and studying performance, which also increased these detection uncertainties. 

\subsubsection{Detector Calibration}

With the neutron beam off, the detection systems were characterized through the use of radioactive sources $^{113}$Sn, $^{207}$Bi, and $^{109}$Cd. 
The radioactive source insertion system used for these measurements was used to calibrate $29\%$ of the total pixels. 
Pixels that did not have enough counts in the peaks to calibrate were instead calibrated by assuming that their gain and nonlinearity had the same mean values as the well-characterized pixels of the detector.
To take into account backscattered electrons in both calibration and neutron-decay data sets, the energy of events arriving within $1$~$\mu$s in active nearest-neighbor or next-nearest-neighbor pixels on the upper and lower detector were summed together. 
Simulations show negligible ($<10^{-5}$) backscattered electrons have a time of flight longer than 1~$\mu$s.
Prominent peaks corresponding to mono-energetic decay electrons in the spectrum of each pixel were fit with a Gaussian added on top of an exponential, which represented electron downscattering, and flat background. 
The data points relating the centroids to the corresponding known electron energies were fit to a polynomial to determine the gain, offset, and nonlinearity of the detector across the range of $\beta$-decay electron energies.
The detector was calibrated across a range of bias voltages and detector temperatures.
On the calibrated pixels, the uncertainty in the electron endpoint was calculated by evaluating the $1$-$\sigma$ uncertainty of this polynomial at $782$~keV. 
This led to a $6.7\times10^{-3}$ relative uncertainty for the energy of the endpoint of the calibrated pixels, dominated by gain variations from the varied run conditions. 
The observed gain distribution on calibrated pixels was used to provide a systematic uncertainty on this calibration technique for the uncalibrated pixels.
The gain distribution on calibrated pixels was fit to a Gaussian with 
a $\sigma$ of $3.1\%$, which was taken as the calibration uncertainty for the uncalibrated pixels. 
The $6.7\times10^{-3}$ uncertainty on the endpoint energy for well calibrated pixels, combined with a $3.1\times10^{-2}$ uncertainty on the endpoint energy of uncalibrated pixels, provides an overall calibration uncertainty at the endpoint energy of $\beta$-decay electrons of $2.5\%$.

\subsubsection{Coincidence events}

In this dataset, waveforms from all triggered hits were stored, so that grouping of coincidences could be studied in offline analysis after applying the energy calibration. 
In analysis, the timestamps of events on the lower detector were shifted by $260$\,ns to account for the synchronization offset between the two DAQ chassis.  
All triggered events depositing an energy less than 50\,keV in the upper detector were flagged as 
possible protons. These protons were grouped with all events that preceded the proton by its expected time of flight of $10$~$\mu$s to $80$~$\mu$s, which were subsequently defined as electrons. 
A spatial cut was applied to separate foreground from background, where foreground events occurred within two pixels of the proton hit, mapped to either the upper or lower detector. 
The rate of coincidence events was $60$\,cps. 

\begin{figure}
    \centering
    \includegraphics[width=\columnwidth]{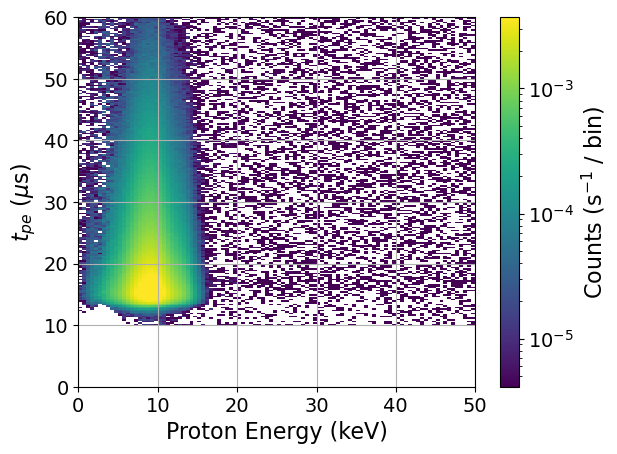}
    \caption{The two-dimensional histogram of the “proton” energy vs. the time of flight between the “proton” and “electron” shown with backgrounds subtracted as discussed in the text.}
    \label{fig:proton_peak}
\end{figure}

Fig.~\ref{fig:proton_peak} shows the time of flight between protons and electrons histogrammed over all coincidences. 
In this dataset, the proton peak appeared at $10$~keV, lower than the amplitude of $20$~keV predicted by simulation and observed using the same detector at the proton source at the University of Manitoba~\cite{Macsai:2023the}. The peak reduction would lead to a bias of several percent in the reconstruction of the main physics observable, the neutrino electron correlation coefficient $a$, due to selectively removing protons with lower momentum.
This was a consequence of accidental contamination after the testing at the University of Manitoba.
The bias in the physics observable depends on the specific deposit properties, which have not been precisely characterized.

\subsubsection{Background rates}

Both silicon detectors are susceptible to unwanted backgrounds originating from electronic noise, cosmic muons, as well as neutron-beam-generated particles. 
Backgrounds in the lower detector were $\sim$$11000$~cps above the DAQ threshold of $\sim$$90$\,keV, significantly higher than the backgrounds in the upper detector, which were only $\sim$$200$~cps above the DAQ threshold of $\sim$$9$\,keV. 
Some of the differences in background rates between the detectors are expected. The upper detector has a much lower sensitivity to beam-induced backgrounds, as the lower detector is much closer to the neutron beam and the shape of the spectrometer magnetic field rejects many charged particles from reaching the upper detector.
The background rate in the lower detector was exacerbated by the much larger than expected electronics noise rate from the leakage current. 
The beam-induced background was measured by varying the magnetic fields and the high voltage applied to the upper detector. 
Real decay events required both magnetic fields and high voltage to detect the proton and electron in coincidence. Beam-induced backgrounds are not expected to have the same coincidence structure, and can thus be isolated. 
A significant fraction of backgrounds in the lower detector were neutron beam related, such as from neutrons interacting with the magnesium window separating the ultra-high vacuum bore from the neutron beamline. 

The long coincidence window, from 10\,$\mu$s up to 80\,$\mu$s, means that background events can form accidental coincidences.
The impact of backgrounds was mitigated by requiring coincidences between the upper and lower detector spatially located within 2 rings of pixels, corresponding to an area of $\sim17$~cm$^2$ for each detector. This corresponds to about $18\%$ of the total active area of the detectors. 
Background events with a similar timing structure to the neutron decay signal of interest were generated by applying the coincidence time criteria using a time of flight window between 80 and 150\,$\mu$s. 
The measured time of flight of these backgrounds were shifted by 70\,$\mu$s. Then these events were subtracted from the coincidence data.
This background timing window was chosen for the ease of having one contiguous coincidence time window, from 10 to 150\,$\mu$s, that could be subdivided into equal parts foreground and background. 
Approximately 2\% of proton-electron coincidences from neutron decay appear between 80 and 150\,$\mu$s, slightly inflating the observed background rate.
Furthermore, protons backscattering off the lower detector would appear at high $t_{pe}$, also contributing to backgrounds. 
The measured background coincidence rate was 20~cps, leading to a background subtracted rate of 40~cps.

The backgrounds ultimately caused systematic uncertainties in neutron decay kinematic reconstruction. 
The background due to false coincidences was roughly constant in $t_{pe}$, but the real proton time of flight distribution was nearly flat in $1/t_{pe}^2$. 
Protons backscattering off the lower detector would have a similar time profile as the decays of interest, but would appear at higher $t_{pe}$.
An incorrect estimation of the background rate would introduce nonlinearities in the slope of the $1/t_{pe}^{2}$ distribution on the Dalitz plot.
Furthermore, electron energy reconstruction was biased as a result of detector backgrounds being mistakenly counted as backscattered electrons. 
This bias was mitigated by choosing a 1~$\mu$s integration window for counting backscattered events. 
Scanning this integration window from $0.5$~$\mu$s to $2$~$\mu$s showed no statistically significant variation in the slope of the Dalitz plot. 

\subsection{Simulated Performance}
\label{ssec:simulation}

The response of the spectrometer has been modeled in a detailed Monte Carlo simulation using Geant4 including an analytical description of the electric and magnetic fields as described in \cite{Fry:2018kvq}. 
$10^{8}$ coincidence decay proton and electron events with momenta following the expected neutron decay distributions, neglecting recoil order corrections, were generated in a cylinder in the center of the decay volume. 
The decay volume used in these simulations was a 
vertical cylinder with radius $3.4$~cm and height $8$~cm aligned along the magnet axis and centered in the magnet $13.2$~cm below the field maximum. A cylindrical geometry was chosen because electrons and protons far from the central magnet axis predominately scatter off internal components of the spectrometer rather than reaching the detectors. 
The simulation tracked decay particles through the spectrometer and recorded their arrival time, energy, and position in each silicon detector past a 100\,nm dead layer. 
For events where multiple backscattered electrons hit on the same pixel, the energy of those hits was summed and the time of those events was averaged. 
Each pixel had a simulated 10\,keV threshold on the upper detector or a 100\,keV threshold on the lower detector. Pixels that were off during the measurement were also turned off in simulation. 
Coincidences were reconstructed by selecting proton events on the upper detector and grouping them with electron events preceding the proton by $10$ to $80$\,$\mu$s. The proton time of flight $t_{pe}$ was defined as the elapsed time between the first electron event and the proton event, and the total electron energy $E_e$ was defined as the sum of energies deposited in all electron events. 

\begin{figure}
    \centering
    \includegraphics[width=\columnwidth]{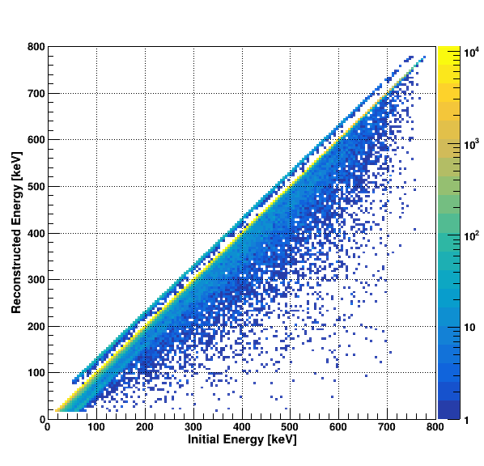}
    \caption{Reconstructed energy as a function of true initial energy for decay electrons generated in a Geant4 simulation of the Nab experiment. (Color online)}
    \label{fig:electronresponse}
\end{figure}

Several factors impacted the energy assigned to the electron compared to its initial energy in the decay.
The electron has a significant energy- and angle-dependent probability to backscatter from the detector~\cite{Martin:2003st}, and all hits must be summed to assemble the full energy. That summation is complicated by the fact that one detector is held at -30\,kV and the difficulty in determining bounce order due to the fast ($\sim50$\,ns) electron travel time between detectors, shorter than the rise time of electron waveforms. If the final electron hit is on the upper detector, $30$~keV must be added to the reconstructed electron energy to account for energy lost in the proton-accelerating potential. 
The electron energy reconstruction is imperfect due to missed backscattered electrons, exacerbated by the missing detector pixels.
The electron energy reconstruction also must account for energy loss due to bremsstrahlung and due to the small fraction of its energy deposited in the inactive dead layer of the silicon detector. 

Finally, the calibration of the detector response must be exceptionally well understood. The simulated detector response is shown in Fig.~\ref{fig:electronresponse}. Most events are reconstructed correctly and fall within the diagonal line where initial and reconstructed energy match. 
Events below this line on the y-axis have experienced electron energy losses as described above, and events above this line on the y-axis have a misidentified bounce order resulting in a 30\,keV shift due to the upper detector high voltage.

\begin{figure}
    \centering
    \includegraphics[width=\columnwidth]{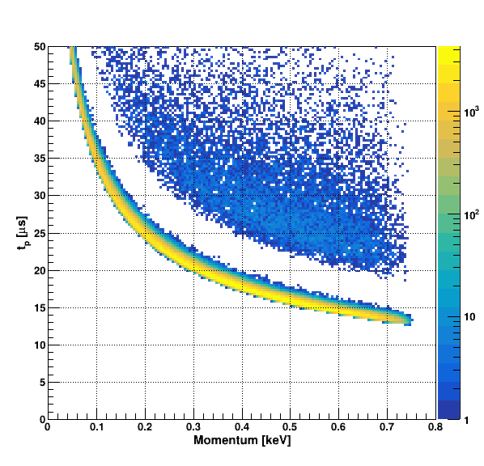}
    \caption{The proton time of flight $t_p$ shown as a function of the initial momentum of the decay proton expected for the Nab spectrometer response simulated in Geant4. (Color online)}
    \label{fig:protonresponse}
\end{figure}

A challenge unique to the Nab experiment due to the spectrometer response is the assignment of proton momentum given the measurement of the proton's time of flight $t_p$. 
Assuming adiabatic transport along the magnetic field, the initial momentum of the proton, $p_p$, can be related to the time of flight of the proton $t_p$ between the initial decay coordinate $z_0$ and position of the upper detector $l$, across the magnetic field region, with \cite{GlueckAspect:2005, Baessler:2014gia}:
\begin{equation}
    t_p = \frac{m_p}{p_p}\int_{z_0}^{l}{\frac{dz}{\sqrt{1-\frac{B(z)}{B_0}\sin^2{\theta_0} + \frac{q \left(V(z) - V_0\right)}{E_0}}}},
    \label{eq:proton_mom}
\end{equation}
where $\theta_0$ is the angle of the initial momentum with respect to the initial magnetic field $B_0$, $V_0$ is the initial electrostatic potential, and $E_0$ is the proton's initial kinetic energy. 
The simulated time of flight of protons as a function of initial proton momentum due to the spectrometer response, cf. Eq.~\ref{eq:proton_mom}, is shown in Fig.~\ref{fig:protonresponse}. 
The length of the spectrometer was measured to $\pm5$\,cm. 
Since protons travel more than $5$\,m, this corresponds to a $<1\%$ uncertainty in the simulated length of travel.
This does not include small biases in the assignment of particle time stamps by the detection system, such as described in Ref.~\cite{Hayen:2022dsg}. The Nab spectrometer magnetic fields were optimized to produce a narrow width in $t_p$ for accepted protons, resulting in the narrow distribution with an inverse proportionality between proton time of flight and momentum. The events at longer time of flight than the main band are protons backscattered from the lower detector to the upper detector. 
For the remainder of this work, we report the decay distribution in terms of $t_{pe}$, which incorporates the electron time of flight,
rather than mapping $t_p$ to $p_p$ using Eq.~\ref{eq:proton_mom}. This intentionally blinds us to the true slope of Dalitz plot, and thus the value of the $a$ coefficient.

\begin{figure*}[ht!]
    \centering
    \includegraphics[width=0.49\linewidth]{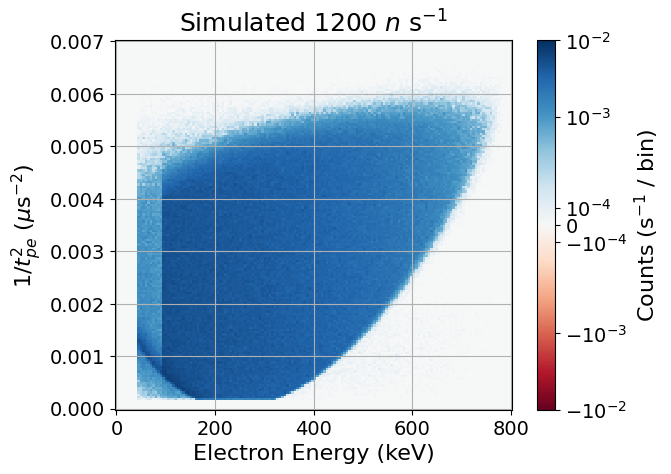}
    \includegraphics[width=0.49\linewidth]{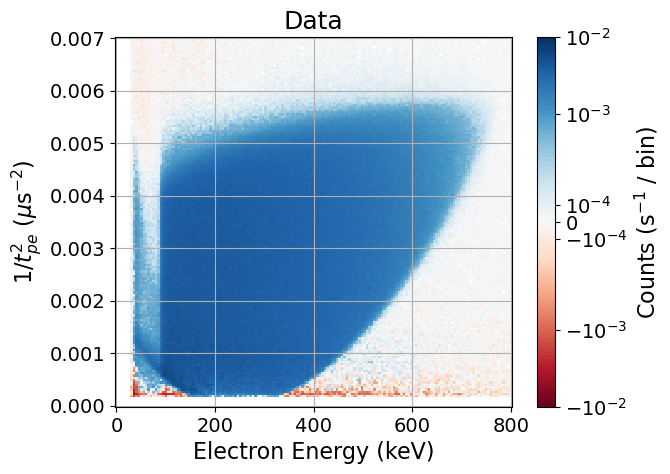}
    \caption{(Left) The Dalitz plot of neutron $\beta$-decay phase space expected for the Nab spectrometer. (Right) The neutron Dalitz plot measured in the Nab apparatus. Details are discussed in the text.}
    \label{fig:teardrop}
\end{figure*}

The coincidence rate in the detector, $R_{det}$, can be used to estimate the neutron decay rate, $R_{tot}$, in the experiment during typical operation with the monochromator~\cite{Fomin:2014hja} inserted:
\begin{equation}
    R_{det} = f_{filt} f_{thresh} f_{pix} R_{tot}\\.
\end{equation}
The counting efficiencies $f_{filt}$, $f_{thresh}$, and $f_{pix}$ can be determined through these Geant4~\cite{geant4} simulations.
The presence of the $4$~T magnetic field above the decay volume allows only $f_{filt} \approx 12\%$ of the protons to reach the upper detector.
Simulated particle detection thresholds of $100$~keV on the lower detector and $10$~keV on the upper detector provide an $f_{thresh} \approx 90\%$. 
The reduction in the total number of available pixels due to hardware failures meant that we saw a reduced number of these protons. By turning off the pixels in simulation which were off in the experiment, we simulate a counting efficiency of $f_{pix} \approx 32\%$. 
Since we observed $R_{det} \approx 40$~cps the simulated proton detection efficiencies allow us to estimate $R_{tot} \approx 1200$ decays\,s$^{-1}$ in the Nab spectrometer decay volume during the commissioning data-taking, averaging over beam conditions. 

In Nab, the neutron decay phase space depicted in Fig.~\ref{fig:intro:kinematics} is related to the experimentally accessible parameters of electron energy and proton time-of-flight as depicted in Figs.~\ref{fig:electronresponse} and ~\ref{fig:protonresponse}. The simulated Dalitz plot as a function the observables of $t_{pe}^{-2}$ and reconstructed electron energy in the Nab apparatus is shown in Fig.~\ref{fig:teardrop} (left). 
The simulation was scaled to a neutron decay rate of 1200 decays s$^{-1}$, and included an applied threshold on the total energy in each pixel of $10$~keV on the upper detector and $100$~keV on the lower detector. Additionally, pixels which were not active during the measurements reported here were excluded from the simulation. 
Data taken at the Manitoba proton source demonstrated that the lower-than-expected proton signal-to-noise observed in the dataset here introduces a $\sim250$~ns timing jitter on proton events as the noise makes finding the rising edge of the proton waveform less precise. 
The simulated response is blurred by randomly adding a gaussian-sampled time to protons with $\sigma=250$~ns to account for this.

\subsection{Results}
\label{ssec:coincidences}

\subsubsection{Neutron \texorpdfstring{$\beta$}{Beta} Decay Dalitz Plot}

The Nab experiment recorded $1.6\times10^7$ background-subtracted neutron decay events over 101 hours of production data-taking, which corresponds to a live-time of 81 hours of data. 
The raw waveforms were convolved offline with a trapezoidal filter with settings chosen from Table 1 of Ref.~\cite{Jezghani:2020}, which tuned its parameters based on data taken in Ref.~\cite{Broussard:2017tab}. 
The energy of events was determined using a ``long trapezoid,'' with a rising edge of $1.6$~$\mu$s and a flat top of $0.4$~$\mu$s. 
The timing of events was determined with a ``short trapezoid,'' with a rising edge of $80$~ns and no flat top. 
Raw waveforms with a rise time - defined as the time for the amplitude going from 10\% to 90\% of the peak - greater than $400$\,ns were considered to be noise 
or non-backscattered-electron pileup 
and rejected. The reconstructed energies and timings were used to generate coincidences as described above. Background subtraction used the shifted time window procedure, also described previously. 
The reconstructed Dalitz plot from the measured dataset can be seen in Fig. \ref{fig:teardrop} (right).

Several features are apparent when comparing the measured and simulated Dalitz plots in Fig.~\ref{fig:teardrop}. Qualitatively, the agreement in the ``teardrop'' shape indicates no severe systematic biases in the measurement.
A nonuniformly reduced count rate in the region between 10\,keV to 100\,keV is apparent in the measured data.
The events in Fig.~\ref{fig:teardrop} between $10$~keV and $100$~keV correspond to both the electron and proton only hitting the upper detector.
The reduction is caused in part by fast protons hitting the same pixel as an electron, such that they are on the tail of the electron pulse and are missed by the analysis routine. The reduced proton trigger efficiency for these events is worsened by the degraded proton energy. 
Another visible difference is evident at small $1/t_{pe}^2$. 
The lower electrode was operated intermittently in the measured dataset presented in Fig.~\ref{fig:teardrop}, which has two competing effects that are poorly characterized. 
For periods where it was not operated, backscattered protons were expected appear at a delayed time which appear as an excess in low $1/t_{pe}^2$. 
Delayed protons also contribute to over-subtraction of backgrounds which produces the deficit at very low $1/t_{pe}^2$ corresponding to $t_{pe}=70-80$\,$\mu$s.

\begin{figure}
    \includegraphics[width=\columnwidth]{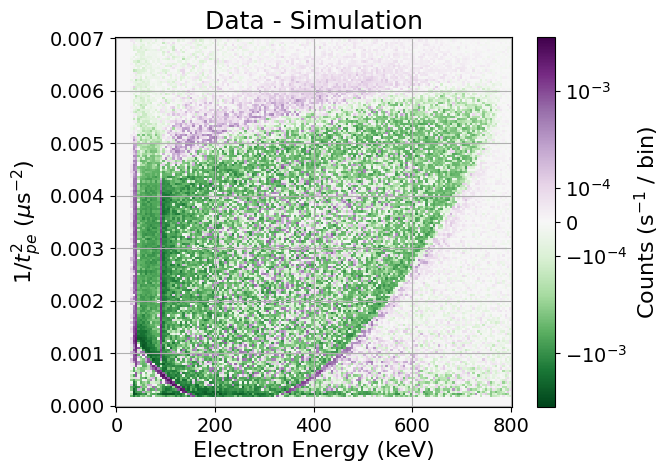}
    \caption{Difference between the Dalitz plots     expected from simulation and from measurement in Fig.~\ref{fig:teardrop}.
    }
    \label{fig:sim_teardrop_diff}
\end{figure}

The difference between the simulation and measurement is shown in Fig. \ref{fig:sim_teardrop_diff}. The simulated teardrop has been scaled to the estimated rate of $1200$ decays s$^{-1}$. 
Below $E_{e} = 200$~keV, the simulation has more counts than the data.
Imperfect reconstruction of the electron energy changes both the central value of the reconstructed electron energy, and broadens the teardrop across the horizontal axis. The effect of electron energy loss, as in Fig.~\ref{fig:electronresponse}, is most noticeable at lower energies. Measured events near the threshold have a variable trigger efficiency, which contribute to a bias in reconstruction of energies in the lower electron energy section of the Dalitz plot. 
Missed backscattered electrons, particularly due to missing pixels, exacerbate this distortion. 
Vertical bands near $10$ and $100$~keV indicate a mismatch between the simulated and real threshold, partially due to the lack of electronic noise in simulation.
The observed discrepancies at low energies also indicate potential inaccuracies in the simulated energy deposition, backscattering probabilities, or backgrounds affecting the energy reconstruction. Furthermore, the impact of the underdepleted lower detector 
on electron energy and timing reconstruction was not well characterized.

Additional $t_{pe}$ blurring (vertical axis) beyond the noise-induced jitter of $250$~ns measured at the University of Manitoba appears to be present in Fig. \ref{fig:sim_teardrop_diff}.
The jitter in extracting the proton timing from a waveform strongly increases as the signal-to-noise decreases. 
Because the incident proton energies were so near the threshold, small variations in the signal-to-noise during the run led to large changes in the jitter of $t_{pe}$. 
A smaller disagreement in $t_{pe}$ can be attributed to uncertainties in the DAQ synchronization between the chassis. 
The simulations used in this result neglect variation in charge collection time, which manifest as a timing bias. 
These have not been quantified but are expected to be smaller than the 5\,ns DAQ synchronization uncertainty~\cite{Hayen:2022dsg}. 

\begin{figure}
    \centering
    \includegraphics[width=\columnwidth]{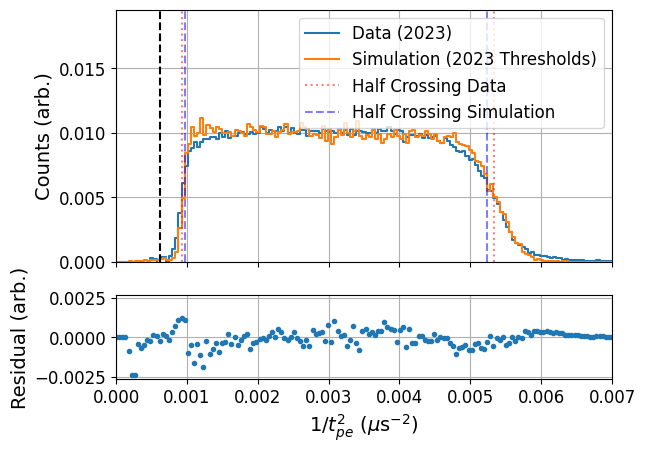}
    \caption{(Top) Histogram of $1/t^2_{pe}$ for background subtracted $p$-$e$ coincidence events with a reconstructed electron kinetic energy of $440$-$460$~keV, for both data and simulation. The histograms are each scaled so that the integral is one. The blue and red vertical lines indicate the midpoint of the edges, and the black vertical line indicates $40$~$\mu$s. (Bottom) Differences between measured data and simulation histograms.}
    \label{fig:trapezium}
\end{figure}

\begin{figure}
    \centering
    \includegraphics[width=\columnwidth]{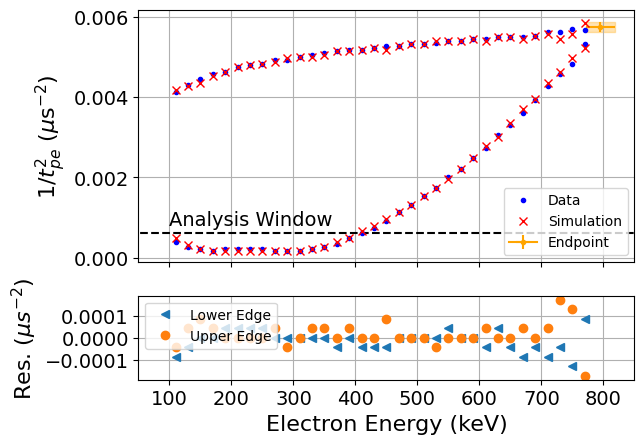}
    \caption{
    (Top) Mid-point of the upper and lower edges of the Dalitz plots in Fig.~\ref{fig:teardrop} as a function of electron energy for measured data and simulation (without uncertainties). 
    The orange point and associated box illustrates the maximum $E_e$ and $p_p$ point as determined in Sec.~\ref{sec:excited_neutrons}, with their corresponding uncertainties. 
    (Bottom) Absolute differences between data and simulation for both the upper and lower edges as a function of electron energy.}
    \label{fig:teardrop_edges}
\end{figure}

To quantify the level of agreement between simulation and measurement, the edges of the measured and simulated Dalitz plots in Fig.~\ref{fig:teardrop} were defined by first binning into 20\,keV wide bins in electron energy $E_e$, and then determining the value of $1/t_{pe}^2$ corresponding to the midpoint in the rising and falling edges for each bin. 
A representative example of one of these energy slices, $440$-$460$\,keV, can be seen in Fig.~\ref{fig:trapezium}.
The resulting midpoints for upper and lower edges of the Dalitz plots for simulation and measurement as a function of electron energy, along with residuals, is shown in Fig.~\ref{fig:teardrop_edges}. The maximum observed fractional deviation between simulation and measurement for the upper edge is $7.8\times10^{-3}$ and for the lower edge is $5.6\times10^{-2}$.
For an extraction of $a$, the slope of the distribution will be fitted between $t_{pe} = 10$\,$\mu$s and $t_{pe} = 40$\,$\mu$s, a narrower timing window than this work. 
The area of greatest discrepancy along the lower edge corresponds to an area outside this intended $a$ analysis window of $40$\,$\mu$s.
Due to the reduced proton energy, the fast protons (the upper edge) are more likely to be counted than slow protons (the lower edge), which biases the count rate at the edges of the Dalitz plot. 
The data set described here is systematically limited primarily by three effects, each on the order several percent: the missing pixels due to hardware and electronics failures which greatly increased the calibration uncertainty; the uncertainty in the proton $t_{pe}$ due to a combination of noise-induced jitter and the DAQ synchronization; and the bias in the teardrop edges due to the degraded proton peak.

\subsubsection{Limits on Excited Neutrons}
\label{sec:excited_neutrons}

\begin{figure}
    \centering
    \includegraphics[width=\columnwidth]{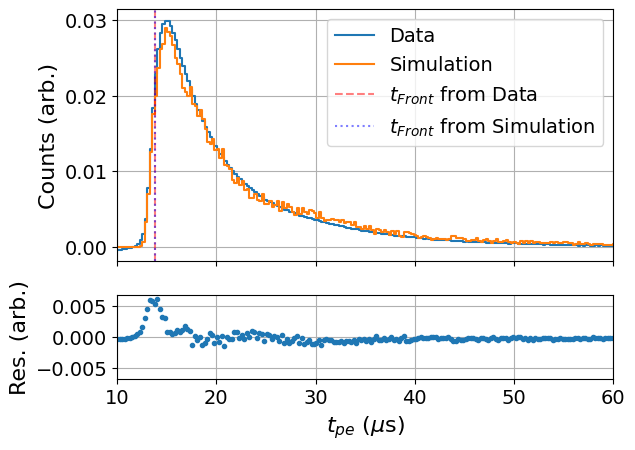}
    \caption{(Top) Distribution of proton times of flight for measured data and simulation, scaled such that the integral from $10$~$\mu$s to $60$~$\mu$s is $1$. (Bottom) Differences between measured data and simulation histograms.}
    \label{fig:proton_tof}
\end{figure}

We now use the observed Dalitz plot from free neutron decay to place a limit on an exotic excited neutron state and decay mode proposed in Ref.~\cite{Koch:2024cfy}. 
The excited neutron was suggested as an explanation for the higher value for the neutron lifetime observed in cold neutron beam experiments compared to ultracold neutron storage experiments \cite{ParticleDataGroup:2024cfk}. In a beam experiment, the protons from decaying neutrons are observed much closer in time to neutron production in a fission or spallation process than in a storage experiment. As a result, a much higher proportion of these decaying neutrons may be in their hypothetical excited state. 
The difference between the two measured neutron lifetimes could be due to a difference in lifetimes between the SM neutron and excited neutron.
The key observable of such an excited neutron would be additional available decay energy or a higher mass than the SM neutron. This energy can be released through the spontaneous emission of a $\gamma$ ray with energy $\Delta E_\gamma$, a process with a characteristic lifetime $\tau_\gamma$. 
Alternatively, this excited neutron can decay directly to a proton, electron, and antineutrino, albeit with extra energy in the decay products.
This extra energy would appear in the Nab spectrometer as either an increase in the maximum proton momentum or an increase in the maximum electron energy beyond the SM prediction. 

The agreement of the maximum $p_p$ with the SM prediction can be determined by comparing the observed $t_{pe}$ in the Nab spectrometer with the simulated $t_{pe}$. 
The $t_{pe}$ distribution was summed over all electron energies to remove the effect of calibration uncertainties. 
This comparison can be seen in Fig.~\ref{fig:proton_tof}.
A shift $\Delta t$ between the measured and simulated $t_{pe}$ distributions can be interpreted as a measurement of excess proton momentum in neutron decay.
The front edge of the $t_{pe}$ distribution, $t_{\text{front}}$, for both measured data and simulation was determined through a fit to an empirical function $f(t_{pe})$: 
\begin{equation}
    f(t_{pe}) = A \frac{e^{-(t_{pe}-t_{\text{front}})/k_1}}{1 + e^{-(t_{pe}-t_{\text{front}})/k_2}} \\,
    \label{eq:proton_tof_fit}
\end{equation}
where $A$ is the number of decays and $k_1$ and $k_2$ characterize the sharpness of the falling and rising components of the function.
The $t_{\text{front}}$ from the data was found to be $17$~ns longer than the simulated $t_{\text{front}}$. 
The fitting uncertainties of both the measured and simulated $t_{\text{front}}$, combined in quadrature with the $5$~ns uncertainty in the DAQ timing synchronization, provide an overall uncertainty on the timing discrepancy between measurement and simulation of $21$~ns. 
To validate the empirical fit function of Eq.~\ref{eq:proton_tof_fit}, the upper limit of the fit was varied from $20$\,$\mu$s to $60$\,$\mu$s. The difference in $t_{\text{front}}$ between measurement and simulation was consistent across this fit range within this $21$~ns uncertainty.

The proton time of flight in simulation depends on the accuracy of the modeled position of the collimators defining the neutron beam position and the position of the ground electrode system surrounding the upper detector. The individual components were physically measured with few mm precision with respect to the spectrometer.  Surveys were also performed using laser tracking of multiple permanent fiducial markers on the spectrometer bore and flanges and optical targets inside vacuum before and after cooling the magnet. The total uncertainty in the length of the spectrometer, taking into account thermal contraction, is conservatively estimated at $5$~cm, as determined from the center of the neutron beam to region where the electric field below the upper detector becomes large.
This corresponds to an uncertainty of $150$~ns, consistent with the agreement observed here. 
Adding a $5$~cm positional uncertainty to account for potential systematics due to the simulation in quadrature with the observed $17 \pm 21$~ns discrepancy leads to an overall shift $\Delta t = 17 \pm 151$~ns. 
Combining this with the $\approx13$~$\mu$s minimum arrival time between protons and electrons corresponds to a $(2 \pm 13)\times10^{-3}$ fractional discrepancy on the momentum of the fastest protons from the simulation.

When applied to the proton momentum at the endpoint $p_{p,\max}$, this observed discrepancy corresponds to a measurement $p_{p,\max} = 1188 \pm 13$~keV\,c$^{-1}$. 
This momentum can be related to the endpoint energy by kinematics, as at the endpoint $p_{p,\max} = -p_{e,\max}$. Since the total available energy in this process at the endpoint is given by $m_n-m_p-m_e$, where the masses of the electron ($m_e=0.511$~MeV) and proton ($m_p=938.272$~MeV) are well known~\cite{ParticleDataGroup:2024cfk}, this measurement of $p_{p,\max}$ is equivalent to a measurement of the neutron mass $m_n = 939.566 \pm 0.012$~MeV. 
As described in Ref.~\cite{Koch:2024cfy}, previous measurements of $m_n$ are of the SM neutron. 
At $90\%$ C.L., our measurement of the neutron mass gives $m_n < 939.565 + 0.021$~MeV, placing a limit on the available extra energy in neutron decay. 

\begin{figure*}
    \centering
    \includegraphics[width=0.49\linewidth]{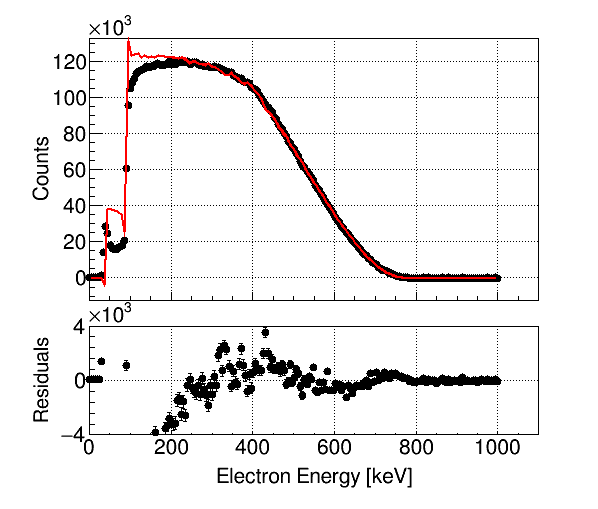} 
    \includegraphics[width=0.49\linewidth]{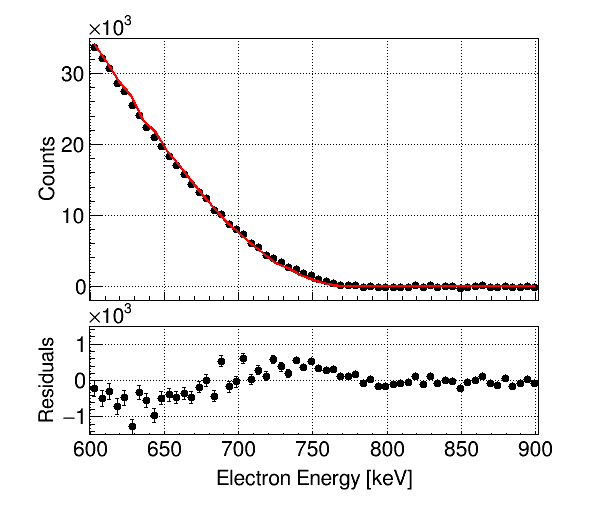}
    \caption{    (Left) Energy spectrum of electron events from neutron decays measured in the Nab spectrometer (black dots) and the fitted simulated response (red line) in the range [200, 1000] keV with absolute residuals below. (Right) Zoom of energy spectrum, fit, and residuals near the endpoint.  Uncertainties are statistical. }
    \label{fig:energy_comp}
\end{figure*}

The available energy in neutron decay was also determined using the measured endpoint of the $\beta$-energy spectrum, with similar precision. We do not characterize the uncertainty in the $\beta$-energy reconstruction due to the underdepleted lower detector and the missing pixels. However, as high-energy electrons are used for calibration and the maximum energy of the $\beta$-spectrum is used to determine the available energy in neutron decay, sensitivity to spectral distortion is reduced. Because of these systematic considerations, we used the $\beta$-spectrum endpoint only as an independent check of the available energy in the decay. 

To find the endpoint energy, the expected spectrum shape from the Geant4 simulation was used as a fit function. An additional cut was applied to keep only coincidences with 12.5\,$\mu$s\,$< t_{pe} <$\,40\,$\mu$s. The simulated energy spectrum was histogrammed with 5\,keV bins and a cubic spline was used for interpolation to obtain a smooth function. Recoil order corrections have a negligible effect on the endpoint and were neglected. The simulation assumed the Fierz interference term $b=0$; a non-zero $b$ would distort the shape of the spectrum but not affect the endpoint. Fig.~\ref{fig:energy_comp} shows a $\chi^2$ fit of the measured data to simulation along with the absolute residuals. The total normalization, energy calibration factor, and constant background were free parameters in the function. The distortion of the spectrum due to imperfect energy reconstruction is evident at low energies below 300\,keV. Above the endpoint the constant background fit parameter accounted for a small over-subtraction of measured background, likely due to the timing window used for estimating the background.

The uncertainty in the determination of the endpoint energy from the fit method was conservatively taken to be $1.5\%$ by varying fit parameters and fit range, in particular the normalization parameter to capture possible impact of missed backscatters. 
Adding in quadrature the $2.5\%$ uncertainty from the calibration, the total systematic uncertainty in the endpoint energy of the decay electron was $2.9\%$.  
The measured endpoint energy from the best fit was found to be $795 \pm 23$~keV, or a measurement of the neutron mass of $m_n=939.578 \pm 0.023$~MeV. This determination is consistent with the approach using the minimum proton time of flight.

\begin{figure}
    \centering
    \includegraphics[width=\columnwidth]{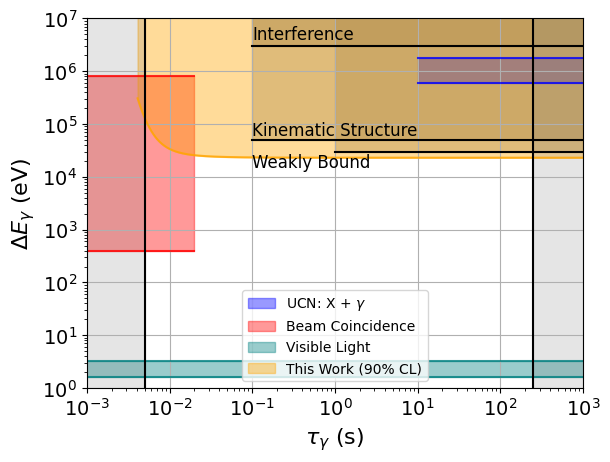} 
    \caption{Available parameter space for the energy of the gamma emitted by excited neutrons, assuming the spallation process produces purely excited neutrons that travel to the Nab spectrometer with a travel time $t_{tr}=22$\,ms.   Other exclusion regions are described in~\cite{Koch:2024cfy}.  The neutron lifetime discrepancy can be explained by excited neutrons for values of $\tau_{\gamma}$ between the vertical grey bands. See also Ref.~\cite{ExcitedBlatnik:2024} for a slightly tighter upper limit for this region. The orange band is the region directly excluded by the measurement of the free neutron mass in $\beta$-decay described in this work. The top end of the plot is the approximate energy released per neutron in a fission or spallation process; if $\Delta E_{\gamma}$ is too high, excited neutrons cannot be generated.  }
    \label{fig:excited_limits}
\end{figure}

If the travel time between neutron generation due to spallation and decay in our spectrometer $t_{tr}$ is much less than the hypothetical de-excitation time $\tau_\gamma$, we will have a sample of almost pure excited neutrons. The converse is true as well; if $t_{tr} \gg \tau_\gamma$, our ability to measure excited neutrons is reduced as most excited neutrons decay to SM neutrons in transit. The number of excited neutrons $N_E(\tau_\gamma)$ and SM neutrons $N_{SM}(\tau_\gamma)$ starting from an initial number $N_0$ can be written as:
\begin{equation}
\begin{split}
    N_E(\tau_\gamma) &= N_0 e^{-t_{tr} / \tau_\gamma} \\
    N_{SM}(\tau_\gamma) &= N_0 \left(1 - e^{-t_{tr} / \tau_\gamma}\right) .
\end{split}
\label{eq:halflife}
\end{equation}
The Nab magnet is installed about $17$~m from the moderator, and the peak flux in the neutron spectrum occurs at about $\sim5$~\AA ($790$~m/s)~\cite{Fomin:2014hja}. This suggests a travel time $t_{tr} = 22$~ms between the neutron generation and decay. The Geant4 simulation of the Nab spectrometer was used to determine the sensitivity of the $t_{pe}$ extraction to the mixing of neutrons and excited neutrons in the decay volume. Two simulations were used: one with only SM neutrons and one with only excited neutrons (with $23$~keV of extra energy). These two simulations were used to find $t_{\text{front}, SM}$ and $t_{\text{front}, E}$ where the subscripts indicate the SM or excited neutron simulations. Intermediate admixtures were studied by scaling the $t_{pe}$ histograms of these two simulations by Eq.~\ref{eq:halflife} for a range of $\tau_\gamma$ using $t_{tr}=22$~ms and added together. The combined histogram was then also fit to Eq.~\ref{eq:proton_tof_fit} to produce a $t_{\text{front}, \tau}$. 
For each $\tau_\gamma$ this produced a scaling factor $S(\tau_\gamma)$:

\begin{equation}
    S(\tau_\gamma) = \frac{t_{\text{front}, \tau} - t_{\text{front}, SM}}{t_{\text{front},E} - t_{\text{front},SM}}.
\end{equation}
This scaling factor was subsequently applied to the measured $\Delta t=17$~ns described above for each sampled $\tau_\gamma$ to account for intermediate admixtures of the possible excited neutron decay. The systematic uncertainty due to the $5$~cm positional uncertainty was added in quadrature with the scaled $\Delta t$ to produce a combined final limit. From this result we can place experimental limits on the proposed excited neutron state as shown in Fig.~\ref{fig:excited_limits}.  The vertical grey bands indicate values of $\tau_\gamma$ that would not explain the neutron lifetime discrepancy. The red, blue, and teal colored bands show direct searches for photons at neutron sources. Our limit uses the measurement of the neutron mass to exclude the available energy of the de-excitation gamma, $\Delta E_\gamma$, as shown. Conservation of energy requires that the photon energy cannot be larger than the measured mass difference. The horizontal grey bands indicate other indirect limits using the neutron mass, as described in~\cite{Koch:2024cfy}.

To check the sensitivity of our $\Delta E_\gamma$ limits to the neutron time of flight, this procedure was repeated for times of flight between $t_{tr}=16.5$\,ms and $t_{tr}=27.5$\,ms, corresponding to a $25\%$ uncertainty in $t_{tr}$. In the region uniquely excluded by this work, the change in the limit in $\Delta E_\gamma$ is not visible in Fig.~\ref{fig:excited_limits}. The position of the left edge of our reported limit shifts inside the red ``beam coincidence'' exclusion region and the left vertical grey band.

On its own, this measurement of $m_n$ is not competitive with independent neutron mass measurements from the binding energy of deuterium~\cite{Kessler:1999221}. However, the measurement described in this work uses free neutrons and would be sensitive to a hypothesized excited neutron mass. This produces more stringent limits on the available energy in neutron decay than those from other measurements described in~\cite{Koch:2024cfy}. 

\section{Conclusions}

The Nab apparatus was developed to precisely probe BSM physics by looking for CKM matrix non-unitarity, scalar and tensor currents, and exotic new physics such as dark couplings. The novel asymmetric spectrometer enables a high precision determination of the electron energy and proton momentum from neutron decay. This work presents the first measurement reconstructing the full Dalitz plot of unpolarized neutron decay above thresholds. The edges of the ``teardrop'' shape indicate agreement with simulated expectation at the several percent level. This result was used to determine directly the mass of the free neutron. While this is not competitive with the traditional method that determines the neutron mass in a nuclear environment and corrects for binding energy, it places a limit on the hypothesized excited state of the neutron, which was proposed as a possible resolution to the neutron lifetime anomaly.

The Dalitz plot produced here contains the kinematic data for all practical methods of extracting the electron-neutrino correlation parameter $a$. 
Subsequent high-quality Nab data sets will allow for a robust suite of benchmarking tests for various experimental techniques, and for studies of radiative corrections without relying on integrated observables~\cite{Gluck:2022ogz}. 

The primary goal of the Nab apparatus is a high precision measurement of the electron-neutrino correlation $a$ in neutron decay.  The statistics collected in this commissioning correspond to a $1.1$\% result, but systematic effects present in the commissioning data set are the major limitation.
This work does not extract a value for this correlation. Disagreements between simulations and the measured data reported here are due primarily to three dominant systematic uncertainties: (1) insufficient calibration data; (2) missing pixels distorting energy reconstruction; and (3) the low proton signal-to-noise and its resultant uncertainty in timing. The use of pristine detectors, redesigns of electronics components, and improved synchronization routines substantially reduced the impact of these effects in follow-up measurements which are not reported here. Beyond the immediate goals to determine $a$ to test CKM non-unitarity and search for a nonzero $b$ not present in the SM, future experiments could potentially exploit a polarized neutron beam to produce independent measurements of $A$ or other neutron decay correlations~\cite{pNAB:2025fer}, or utilize the Nab spectrometer to search for decay correlations in nuclear systems. The common goal is to build a more robust dataset for the CKM unitarity test and searches for new physics.

\section{Acknowledgements}

This work is dedicated to the memory of J. David Bowman, who passed away in 2024, and Yu Qian, who passed away in 2019. David was the originator of the idea for the Nab experiment and through his storied career made many contributions to fundamental nuclear physics.
Yu was a graduate student whose career was tragically cut short.

This work was supported through the US Department of 
Energy (DOE) (Contracts DE-AC05-00OR22725 including KB0401072 and Early Career Research Program Awards KB0401022 and KC0402010, DE-SC0019309, DE-SC0008107, DE-SC0014622, DE-FG02-03ER41258, DE-FG02-97ER41042, and 89233218CNA000001 under proposal LANLEEDM), through the National Science Foundation (NSF) (Contracts PHY-2213411, PHY-2412846, PHY-2209590, PHY-0855584, PHY-2412782, PHY-2111363, and PHY-1126683), by the Natural Sciences and Engineering Research Council of Canada (NSERC) (Contracts NSERC-SAPPJ-2019-00043 and NSERC-SAPPJ-2022-00024), and in part by the US Department of Energy, Office of Science, Office of Workforce Development for Teachers and Scientists (WDTS) Graduate Student Research (SCGSR) program, and the Science Undergraduate Laboratory Internship (SULI) program. 
This research was supported through research cyberinfrastructure resources and services provided by the Partnership for an Advanced Computing Environment (PACE) at the Georgia Institute of Technology, Atlanta, Georgia, USA.
This research used resources at the Spallation Neutron Source, a DOE Office of Science User Facility operated by the Oak Ridge National Laboratory. 
We thank K. Borah for useful discussions.

\bibliography{references.bib}
\end{document}